# Why Do Eight Units of Methylammonium Enclose PbI$_6^{4-}$ Octahedron in Large-Scale Crystals of Methylammonium Lead Iodide Perovskite Solar Cell? An Answer from First-Principles Study

Pradeep R. Varadwaj,[a,b,*] Arpita Varadwaj,[a,b] Koichi Yamashita[a,b]

[a]Department of Chemical System Engineering, School of Engineering, The University of Tokyo
7-3-1, Hongo, Bunkyo-ku, Japan 113-8656
[b]CREST-JST, 7 Gobancho, Chiyoda-ku, Tokyo, Japan 102-0076

Methylammonium lead triiodide (CH$_3$NH$_3$PbI$_3$) perovskite solar cell is a gem in the list of photovoltaic semiconductors. Although there are numerous fundamental and technological questions yet to be addressed covering various aspects of this system for its commercialization, this study has employed first-principles DFT to model the [PbI$_6$(CH$_3$NH$_3$)$_n$]$^m$ zero-dimensional nanoclusters. Using the calculated binding energy landscapes, it has answered the question: why the corner-sharing PbI$_6^{4-}$ octahedron is surrounded by eight units of the organic cation in the large-scale supramolecular structures of the CH$_3$NH$_3$PbI$_3$ system in 3D? The synergistic effect of the methylammonium, as well as the consequence of positive and negative cooperative effects associated with intermolecular hydrogen bonding on the supramolecular evolution of the CH$_3$NH$_3$PbI$_3$ crystals is briefly outlined.

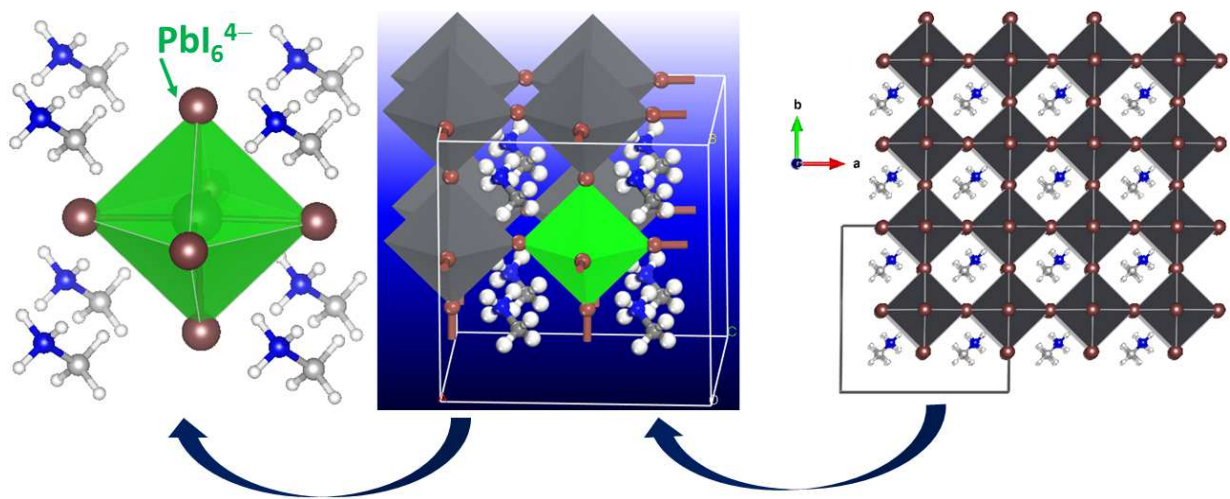

[*] Corresponding Author's E-mail Addresses: pradeep@t.okayama-u.ac.jp (PRV);
varadwaj.arpita@gmail.com (AV); yamasita@chemsys.t.u-tokyo.ac.jp (KY)

Methylammonium lead/tin triiodide ($CH_3NH_3PbI_3$/$CH_3NH_3SnI_3$) perovskite solar cells are the center of many on-going research studies in the area of photovoltaics, both experimentally[1-2] and theoretically using first-principles methods.[3-4] These materials are discovered to serve as efficient semiconductors,[5] especially in converting photonic energy into electric energy.[6]

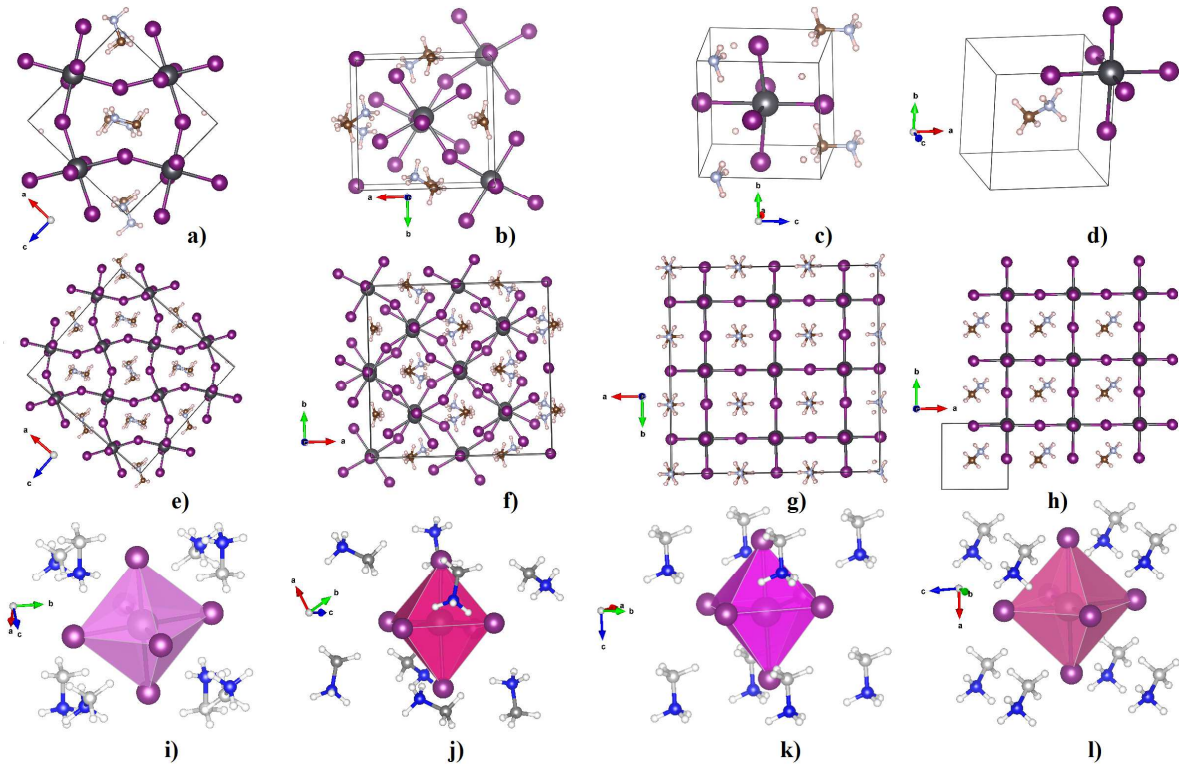

Fig. 1: The unit cells for the a) orthorhombic (*Pnma*), b) tetragonal ($I_4$/mcm), c) pseudocubic (*Pm3m*) and d) pseudocubic (*Pm3m*) phases of the $CH_3NH_3PbI_3$ perovskite solar cell. The e) 2×2×2, f) 2×2×2, g) 3×3×3 and h) 2×2×2 slab models are constructed from the corresponding unit cells, respectively. Shown from i) to l) are the $PbI_6^{4-}$ octahedra responsible for the large-scale emergence of the corresponding slab surfaces, in which, each face of the octahedron is occupied with a single unit of the MA species.

Figs. 1a-d present (top panel) the unit cell geometries for the most important phases of the $CH_3NH_3PbI_3$ perovskite solar cell system. The geometries corresponding to the first three are experimentally identified.[7] These are the so-called orthorhombic (T < 162.4 K), tetragonal (162 < T < 327 K) and pseudocubic (> 327 K), and are evolved through the first and second order phase transitions.[7-8] The geometry of the last one shown in Fig. 1d) has been modelled theoretically using

first-principles methods.[9] It is analogous with the psedocubic geometry presented in Fig. 1c). The main difference between the two pseudocubic geometries lies in the directional orientation of the organic cation.

When the unit cells displayed in Figs. 1a-d are periodically expanded, a variety of slab surfaces can be emerged.[9-10] Figs. 1e-h present, as examples, a kind of such surfaces in 3D. Each of these conceives the corner sharing $PbI_6^{4-}$ octahedron, which is presumably responsible not only for the large-scale grow of the $CH_3NH_3PbI_3$ crystals, but also for making this material and its other derivatives excellent for device application.[11] For instance, it is found that the bandgap of the $CH_3NH_3PbI_3$ solar cell in the tetragonal phase to be 1.61 eV when the Pb–I–Pb angle is quasi-linear, and is ca. 1.65 and 1.69 eV for the orthorhombic and pseudocubic phases, respectively, in which cases, this angle associated with the $PbI_6^{4-}$ octahedra is largely tilted/linear (~160.0° for orthorhombic, and ~180.0° for pseudocubic).[12]

Figs. 1i)-l) illustrate how each of the four $PbI_6^{4-}$ polyhedra in the four respective slab geometries in Figs. 1e-h is topologically enclosed by the eight units of the organic cation. A close inspection of each of these geometries suggests that each triangular $I_3$ face of the $PbI_6^{4-}$ polyhedron is occupied with a single unit the $CH_3NH_3^+$ organic cation in the closed packed solid state structure. It also suggests the orientation of the $CH_3NH_3^+$ cation around the $Pb^{2+}$ core in the orthorhombic phase is quite different from those of the tetragonal and pseudocubic phases.[7-8] Even so, these each is responsible for the development of the aforementioned specific electronic features associated with each of the three phases. An immediate question arises: why does the $PbI_6^{4-}$ octahedron enclose only eight units of the $CH_3NH_3^+$ organic cation, but not more, or not less, than this number? What physical principle is operating on the physical development of the large-scale

supramolecular architecture of $CH_3NH_3PbI_3$ perovskite system that does not allow the occupancy of more or less than eight organic cations around the first coordination sphere of the $Pb^{2+}$ core?

We answer this question by employing density functional theory to the $[PbI_6(CH_3NH_3)_n]^m$ zero-dimensional cluster series, and by the calculating the binding energies of these clusters, where n =1–8 and m is the overall oxidation state of the entire 0D cluster to be determined by the value of n. For instance, when n = 2, m = –2, since the fragments $PbI_6^{4-}$ and $CH_3NH_3^+$ have formal charges of –4 and +1, respectively. Similarly, when n = 4, m = 0; when n = 8, m = +4; and so on. Nevertheless, a standard all-electron DZP basis set retrieved from the EMSL basis set exchange library database was chosen for computation since there are only a few such all-electron basis sets viable in the database for heavy elements like Pb and I.[13-14]

Fig. 2 illustrates the quantum theory of atoms in molecules (QTAIM)[15] based molecular graphs for the $[PbI_6(CH_3NH_3)_n]^m$ clusters, obtained using the PBEPBE/DZP level energy-minimized geometries of the corresponding clusters with Gaussian 09 code.[16] Note that one generally expects the number of clusters for this above series to be eight as n varies from 1 to 8. However, our attempt encountered frequent and abnormal convergence errors during geometry minimizations, thus we did not include into the list the results of the $[PbI_6(MA)_1]^{3-}$ and $[PbI_6(MA)_2]^{1-}$ clusters.

Fig. 2 also summarizes the binding energies ΔE for all the six clusters, including $[PbI_6(MA)_2]^{2-}$, $[PbI_6(MA)_4]$, $[PbI_6(MA)_5]^{1+}$, $[PbI_6(MA)_6]^{2+}$, $[PbI_6(MA)_7]^{3+}$ and $[PbI_6(MA)_8]^{4+}$. These are calculated using Eqn. 1, where E on the right hand side of this Eqn represents to the total electronic energy associated with the energy-minimized geometry of either the cluster, or the isolated monomer.

$$\Delta E([PbI_6(CH_3NH_3)_n]^m) = E(([PbI_6(CH_3NH_3)_n]^m) - E[PbI_6^{4-}] - n \times E[CH_3NH_3^+] \ldots\ldots\ldots 1$$

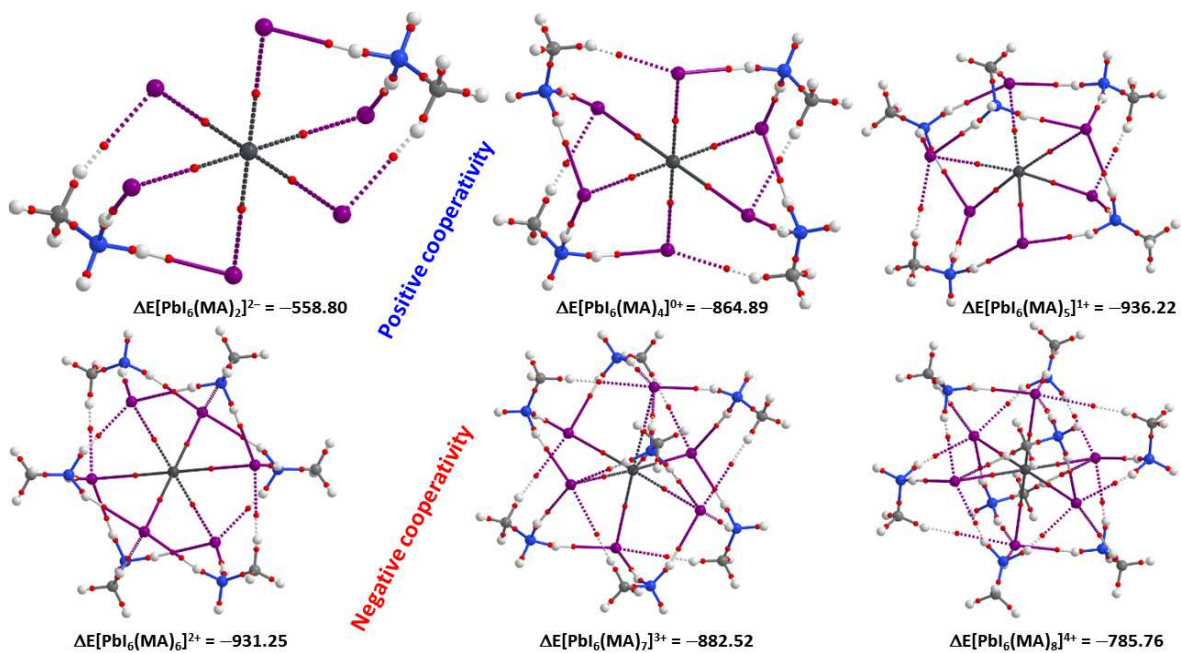

Fig. 2: QTAIM[15] based molecular graphs for the $[PbI_6(MA)_2]^{2-}$, $[PbI_6(MA)_4]$, $[PbI_6(MA)_5]^{1+}$, $[PbI_6(MA)_6]^{2+}$, $[PbI_6(MA)_7]^{3+}$ and $[PbI_6(MA)_8]^{4+}$ clusters in 0D. The binding between MAs and $PbI_6^{4-}$ in each of the three clusters in the upper panel is organized by the principles of positive cooperativity, while of each the three clusters in the lower panel is organized by negative cooperativity. The spherical tiny dots in red, and the solid and dotted lines in atom color between the atomic basins are bond critical points and both paths, respectively. The latter between MA and $PbI_6^{4-}$ in these clusters are reminiscent of intermolecular hydrogen bonding interactions. The binding energies ($\Delta E$) are given in units of kcal mol$^{-1}$.

As such, the $\Delta E$ for the $[PbI_6(MA)_2]^{2-}$ anionic cluster containing two organic cations is calculated to be as large as –558.0 kal mol$^{-1}$. When the number of the organic cations in the $[PbI_6(CH_3NH_3)_n]^m$ cluster is increased to 4, the $\Delta E$ for the resulting cluster $[PbI_6(CH_3NH_3)_4]^0$, which is neutral, became –864.89 kcal mol$^{-1}$. Adding another unit of the organic cation to the surrounding of the $PbI_6^{4-}$ core has further increased the overall size of the modified $[PbI_6(MA)_5]^{1+}$ cluster, with the $\Delta E$ increased to a value of –936.22 kcal mol$^{-1}$. Evidently, these results unequivocally suggest that passing from $[PbI_6(MA)_2]^{2-}$ through $[PbI_6(CH_3NH_3)_4]^0$ to $[PbI_6(MA)_5]^{1+}$, the $\Delta E$ is increased tremendously, revealing the emergence of the phenomenon of non-additive cooperativity, presenting synergistic binding. Note that in each of these latter two cases, the ΔE is substantially larger compared to that of the PbI$_6$(MA)$_2$]$^{2-}$, and the increase in ΔE is positively in each successive step. This suggests that the coupling between the various intermolecular interactions formed between the organic species and that between the organic and inorganic moieties leads to positive cooperativity as it favors and stabilizes the clusters.[17-19]

It is now expected that the ΔE for the successive [PbI$_6$(MA)$_6$]$^{2+}$ cluster will increase largely as the number $n$ of the organic cation in the [PbI$_6$(CH$_3$NH$_3$)$_n$]$^m$ cluster increases to 6. This is, however, not the occasion. Instead, we identified that the ΔE for the [PbI$_6$(MA)$_6$]$^{2+}$ cluster containing six units of the MA cation compared with that of the [PbI$_6$(MA)$_5$]$^{1+}$ cluster containing five units of the same cation to be decreased, with the ΔE for the former ≈ –931.25 kcal mol$^{-1}$. The same feature is noticeable of the ΔE data of Fig 2 upon passing from [PbI$_6$(MA)$_6$]$^{2+}$ through [PbI$_6$(MA)$_7$]$^{3+}$ to [PbI$_6$(MA)$_8$]$^{4+}$. Clearly, while the ΔE for the latter three clusters emanate with the same principle of non-additive cooperativity, the coupling between the interacting species in these clusters disfavors the interaction as the magnitude of the ΔE overall increases. It is what the so-called negative cooperativity is.[17-19] It evolves in the latter three cluster systems with n = 6–8 because as the number of the organic cations go beyond five the steric crowding around the first coordination sphere of the Pb$^{2+}$ core increases significantly. As a result, the intermolecular hydrogen bonding interactions between PbI$_6^{4-}$ and CH$_3$NH$_3^+$ is destabilized, causing the organic cations to be pushed to reside at compromised positions around the first coordination sphere of the Pb$^{2+}$ ion core to maximize the intermolecular interaction between them. Apparently, this is the reason the intermolecular hydrogen bonding distances in these latter three clusters are calculated to be larger than those found for the former three clusters, [PbI$_6$(MA)$_2$]$^{2-}$, [PbI$_6$(MA)$_4$] and [PbI$_6$(MA)$_5$]$^{1+}$.

Clearly, the above results show that there is no space left with the polyhedron that can accommodate a ninth organic cation around it, thus answering the question why the corner-sharing $PbI_6^{4-}$ octahedron does enclose only eight units of the $CH_3NH_3^+$ cation, but not more than this in the MAPbI$_3$ crystals. By contrast, if the $PbI_6^{4-}$ anion accommodates only seven units of the organic cation around it, one on each octant, then an octant of it lacking the organic cation will experience significant repulsive interaction with an identical octant of the nearest $PbI_6^{4-}$ polyhedron facing it. This would result in the complete destabilization of the geometry of the MAPbI$_3$ crystal.

In summary, this study has carried out PBEPBE level density functional theory calculations to insight for the first time into the local equilibrium geometries of the $[PbI_6(CH_3NH_3)_n]^m$ clusters in zero-dimension. Using these, we calculated and analyzed the sequential binding energies, which have significantly assisted us to realize the principles of positive and negative cooperativity in these clusters that have played the leading roles to decide what would be the number of the organic cations around first coordination sphere of the $Pb^{2+}$ ion core. The monomeric fragments are found to interact with each other synergistically through multiple dipole-dipole interactions, leading to the formations of both I•••H–N and I•••H–C intermolecular hydrogen bonding interactions. The results showed that the $[PbI_6(CH_3NH_3)_n]^m$ clusters are ideal prototypes for the fundamental understanding of positive and negative cooperative effects in hydrogen bonding, which have significant implication on the development of materials for supramolecular chemistry.